\documentclass[a4paper,10pt]{article}
\usepackage{authblk}
\usepackage[a4paper,top=3cm,bottom=3cm,left=3cm,right=3cm]{geometry}
\usepackage[T1]{fontenc}
\usepackage[latin1]{inputenc}
\usepackage{ulem}
\usepackage{graphicx,graphics}
\usepackage{amsmath,amssymb,amsfonts}
\usepackage{latexsym,verbatim}
\usepackage{bm,bbm}
\usepackage{dcolumn}
\usepackage{color}
\usepackage{placeins}

\usepackage{color}
\usepackage{subfigure}

\usepackage[percent]{overpic}

\newcommand{\br}{{\bm r}}
\newcommand{\bR}{{\bm R}}
\newcommand{\vD}{v_{\rm D}}

\DeclareMathAlphabet\mathbfcal{OMS}{cmsy}{b}{n}

\begin{document}

\title{Low-frequency critical current noise in graphene Josephson junctions in the open-circuit gate voltage limit}
\date{} 


\author[1,2]{Francesco M. D. Pellegrino}
\author[1,2,3]{Giuseppe Falci}
\author[1,2,3]{Elisabetta Paladino\thanks{elisabetta.paladino@dfa.unict.it}}

\affil[1]{Dipartimento di Fisica e Astronomia ``Ettore Majorana'', Universit\`a di Catania, Via S. Sofia 64, I-95123 Catania,~Italy.}
\affil[2]{INFN, Sez. Catania, I-95123 Catania,~Italy.}
\affil[3]{CNR-IMM, Via S. Sofia 64, I-95123 Catania,~Italy.}

\maketitle 
 
\abstract{
We investigate  critical current noise in short ballistic graphene Josephson junctions 
in the open-circuit gate-voltage limit within the McWorther model. We find flicker noise in a wide frequency range and discuss the
temperature dependence of the noise amplitude as a function of the doping level. At the charge neutrality
point we find a singular temperature dependence $T^{-3}$, strikingly different from the linear dependence
expected for short ballistic graphene Josephson junctions under fixed gate voltage.
} 
\section{Introduction}
\label{intro}

The encapsulation of a graphene sample in hexagonal boron nitride allows the production of devices characterized by
ballistic  transport features up to room temperature at a micrometer scale for a wide range of carrier concentration~\cite{dean_natnano_2010,mayorov_nanolett_2011,wang_science_2013}.
Based on this technology, ballistic bipolar graphene Josephson junctions (GJJ) have been realized~\cite{borzenets_prl_2016,nanda_nano_2017,park_prl_2018}.
These high-quality graphene samples with superconducting electrodes in close proximity to the graphene layer
with ultra-clean interfaces allow Cooper pairs ballistically roaming over micron scale lengths~\cite{benshalom_natph_2015,calado_natnano_2015,english_prb_2016}.
In a GJJ, a dissipationless supercurrent flows in equilibrium through the normal region via the Andreev reflections at each superconductor-graphene interface. 
In this system, the Andreev level spectrum, and as a consequence the current-phase relation, depends on the normal phase channel length ($L$) and 
on the doping level in graphene layer. 
In this work, we focus on the short channel limit, $L \ll \xi$, where $\xi$ is the coherence length in the superconductor.
This regime has been recently experimentally achieved in ~\cite{park_prl_2018}, 
and well described within the  Dirac-Bogoliubov-de Gennes approach~\cite{titov_prb_2006}.
The tunability of the current-phase relation by varying the doping level has allowed the realization of  voltage-controlled transmon, also known as a gatemon~\cite{wang_natnanotech_2018,pellegrino_proceedings_2019}.

In this manuscript, we investigate the critical current noise in short ballistic GJJ within the McWorther model~\cite{mcwhorther_1957} commonly used to describe  noise induced by electron traps in oxide substrates~\cite{hooge_rps_1981,balandin_natnano_2013,paladino_rmp_2014,darrigo_2008,paladino_2010}.
Charge traps act as independent generation-recombination centers, which are described as random telegraph processes.
Due to  proximitized superconductivity of the normal metal forming the junction, fluctuations of carrier density in the graphene insert induce fluctuations of Andreev  levels, resulting in noise  in the critical current of 
the ballistic GJJ \cite{pellegrino_commphys_2020}.

Here, we address the open-circuit gate-voltage limit, excluding any charge flow between the graphene layer and the 
metal gate via the external circuit.
This regime is complementary to the fixed graphene-to-metal-gate voltage-drop 
regime where charge flow through the circuit allows readjustment of the number of carriers in graphene
after each trapping/recombination process. In Ref.~\cite{pellegrino_commphys_2020} the fixed-$V_{\rm G}$ regime was considered under the assumption of instantaneous equilibration of carriers in graphene after each tunneling process. In an equivalent circuit description, in the present manuscript we address the infinite external resistance limit of the bias circuit, 
whereas the zero resistance limit was considered in Ref.~\cite{pellegrino_commphys_2020}. In the following, we will refer 
to these two regimes as the "open-circuit" and "fixed gate voltage" operating conditions, respectively.
We demonstrate that, similarly to the fixed gate voltage regime,
noise in the critical current due to fluctuations of the carrier density in graphene
depends on frequency, $\omega$, as $S_{I_c}(\omega) = \mathcal{A}(T, \mu)/\omega$, where $\mu$ is the chemical potential in graphene. 
For large doping we find that $\mathcal{A}(T, \mu)$ is independent on the operating regime, 
or equivalently on the equilibration time in graphene, as expected.
At the charge neutrality point (CNP) instead we find qualitative differences leading to 
striking consequences in the low-temperature behavior of the noise amplitude $\mathcal{A}(T, \mu=0)$. 
We find that in the open-circuit limit  the noise amplitude diverges  as  $ \mathcal{A}(T, 0)  \propto T^{-3}$, whereas for fixed gate voltage  it
is  $\mathcal{A}_{V_{\rm G}}(T, 0) \propto T$.

\section{Model}
\label{sec:1}

\begin{figure}
\centering
\includegraphics[width=.75\columnwidth]{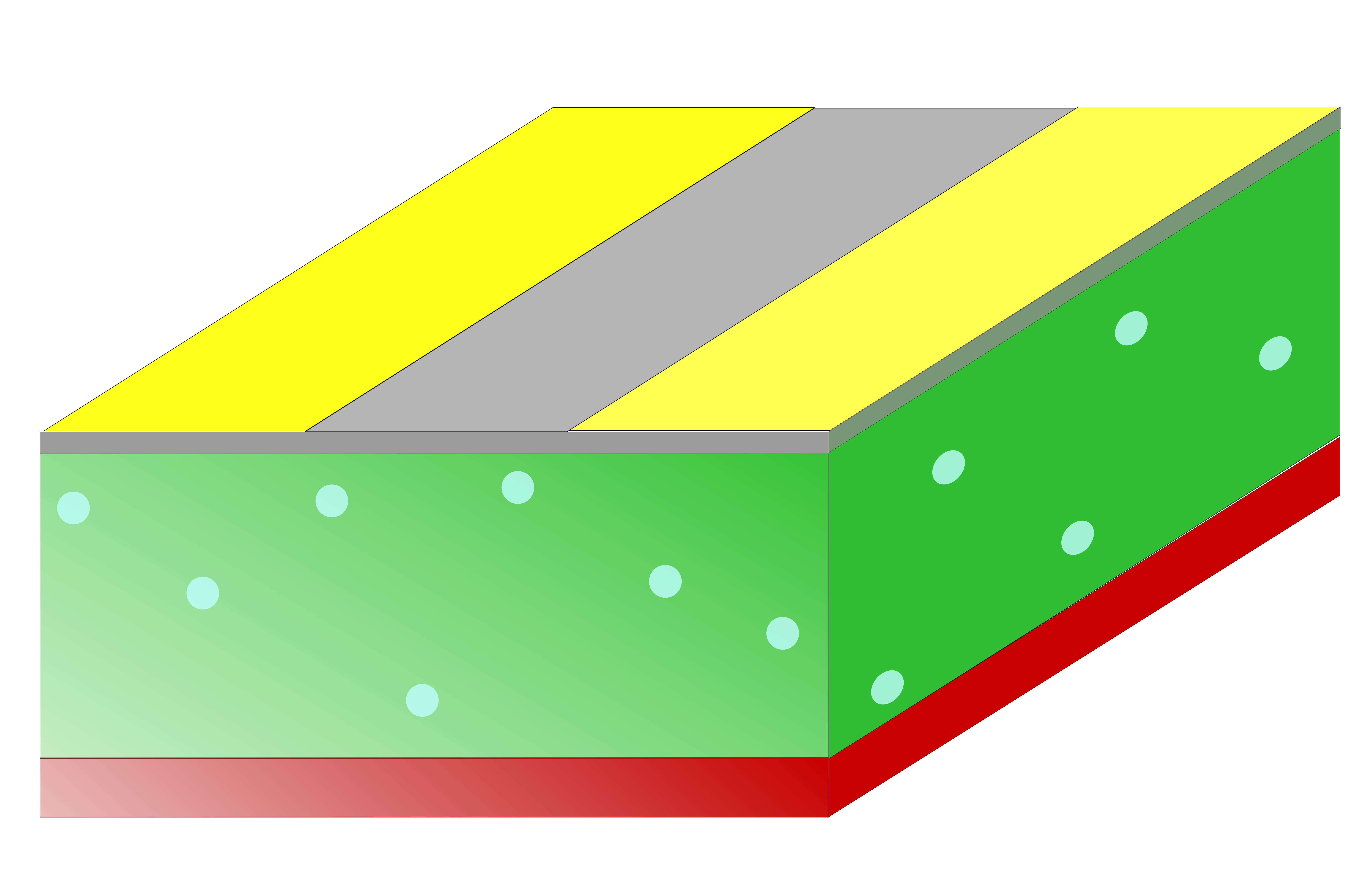}
\caption{Schematic of the device (side view). The device consists of a metal gate
(red), a substrate (green), a monolayer graphene (gray), and two superconducting electrodes (yellow). 
Within the substrate, the cyan circles represent the electron traps. 
}
\label{fig:setup}
\end{figure}

Here we introduce the model which relates the charge carrier density in graphene to charge traps in the substrate.
We refer to the simple device shown in Fig.~\ref{fig:setup}, 
formed by a monolayer graphene on a substrate hosting electron traps placed on top of a metal gate. 
The metal-substrate interface is at $z=0$, the graphene layer is treated as a two-dimensional system at $z=d$ and it is partially covered by two superconducting electrodes. 
The substrate width $d$ is much smaller than both longitudinal sizes along $\hat{x}$ and $\hat{y}$  directions.

Carrier density fluctuations are due to  charge trapping and release processes between graphene  and carrier traps in the underlying substrate, 
and the trapping and recombinations are considered as discrete Markov processes~\cite{kogan_book}.
The occupancy number $X(i,t)$ of the trap labeled by the index $i$ at time $t$ is a random variable. 
Each trap can be empty ($X(i,t)=0$) or occupied by a single electron ($X(i,t)=1$), 
and it randomly switches between these states with time-independent rates (stationary process).
The conditional probability that the trap $i$ at time $t$ has the occupation number $X(i,t)$  if 
the trap $j$ at time $t_0$ has the occupation number $X(j,t_0)$ is written as
\begin{equation}\label{eq:conditional}
 P[X(i,t)|X(j,t_0)]= \delta_{i, j} p_i[X(i, t-t_0)|X(i, 0)] + w_i (X(i, 0))  ~,  
\end{equation}
where $w_i(X(i, 0))$ is the stationary probability of trap $i$ which depends on the  initial occupation $X(i, 0)$
as follows $w_i(1)=f_{i}$ and $w_i(0)=1-f_{i}$, and the Kronecker delta appears because different traps are uncorrelated. 
The matrix $p_i$ is expressed as~\cite{pellegrino_commphys_2020,kogan_book,pellegrino_jstat_2019}
\begin{equation}
 p_i[X(i, t)|X(i, 0)] =
\begin{bmatrix}
 p_i[0|0](t) &  p_i[0|1](t)  \\
 p_i[1|0](t) &  p_i[1|1](t)  \\
\end{bmatrix}=
\begin{bmatrix}
f_i & -(1-f_i) \\
- f_i   & 1-f_i  
\end{bmatrix} e^{-\gamma_i t}~,
\end{equation}
where $\gamma_i=\lambda_{i,00} +\lambda_{i,11}$ is the  switching rate between the two states
of the stochastic process expressed in terms of  the transition rates for  $0 \to 1 $ and $1 \to 0$ processes, 
$\lambda_{i,00}$ and $\lambda_{i,11}$, respectively.
Due to Markovianity, the multi-time correlators reduce to two-points correlation function
\begin{equation}\label{eq:condchain}
 P[X(i_N, t_N)|X(i_{N-1}, t_{N-1});\ldots ;X(i_0, t_0)]=  P[X(i_N, t_N)|X(i_{N-1}, t_{N-1})]~.
\end{equation}
The density of populated traps per unit volume and energy, ${\cal N}_{\rm T}(\epsilon,\bm R,t)$,
fluctuates around its average value ${\cal N}_{{\rm T}0}$.
Assuming that trap $i$ is located at position $\bR_i$ and that the energy of the occupied trap is $\epsilon_i$ (evaluated with respect to the CNP), 
the average value can be expressed as
\begin{equation} 
 {\cal N}_{{\rm T}0}(\epsilon,\bm R)=\sum_{i=1}^{M_{\rm T}} \delta(\epsilon-\epsilon_i) \delta(\bR-\bR_i) f_i~,
\end{equation}
and the fluctuations as
\begin{equation}\label{eq:nt_random}
\delta {\cal N}_{\rm T}(\epsilon,\bR,t )={\cal N}_{\rm T}(\epsilon,\bm R,t)-  {\cal N}_{{\rm T}0}(\epsilon,\bm R)=\sum_{i=1}^{M_{\rm T}} \delta(\epsilon-\epsilon_i) \delta(\bR-\bR_i) [X(i,t)- f_i]~,
\end{equation}
where $M_{\rm T}$ is the total amount of traps. 
From here on, we assume that the stationary probability coincides with the equilibrium occupation function $f_i=f(\epsilon_i,\bR_i)=f_{\rm D}(\epsilon_i-\mu_0)$,
where $f_{\rm D}(x)=1/\{1+\exp{[x/(k_{\rm B} T)]}\}$ is the Fermi-Dirac distribution function and $\mu_0$ is the Fermi level.
Under these conditions, the  average density of populated traps per unit volume and energy and the multi-time correlators read 
\begin{equation}\label{eq:n}
 \langle \delta  {\cal N}_{\rm T}(\epsilon,\bR,t ) \rangle = 0~,
\end{equation}
\begin{align}\label{eq:nn}
& \langle \delta {\cal N}_{\rm T}(\epsilon_1,\bR_1,t_1 )\delta {\cal N}_{\rm T}(\epsilon_0,\bR_0,t_0 ) \rangle = 
\\
& \delta(\bR_1 -\bR_0) \delta(\epsilon_1-\epsilon_0)  {\cal D}(\epsilon_0,\bR_0) 
f_{\rm D}(\epsilon_0- \mu_0)[1-f_{\rm D}(\epsilon_0- \mu_0)]\exp[-\gamma(\epsilon_0, \bR_0)(t_1-t_0)]
~,\nonumber
\end{align}
\begin{align}\label{eq:nnn}
 &\langle \delta {\cal N}_{\rm T}(\epsilon_2,\bR_2,t_2 ) \delta {\cal N}_{\rm T}(\epsilon_1,\bR_1,t_1 ) \delta {\cal N}_{\rm T}(\epsilon_0,\bR_0,t_0 ) \rangle = 
\delta(\bR_2-\bR_1)\delta(\bR_1 -\bR_0)  \\
 &\times \delta(\epsilon_2-\epsilon_1)\delta(\epsilon_1-\epsilon_0) 
 {\cal D}(\epsilon_0,\bR_0) ) f_{\rm D}(\epsilon_0- \mu_0)
[1-f_{\rm D}(\epsilon_0- \mu_0)]  [1-2 f_{\rm D}(\epsilon_0- \mu_0)]  \nonumber \\
 &\times \exp[-\gamma(\epsilon_0,\bR_0)(t_2-t_0)]~, \nonumber
\end{align}
\begin{align}\label{eq:nnnn}
& \langle \delta {\cal N}_{\rm T}(\epsilon_3,\bR_3,t_3 ) \delta {\cal N}_{\rm T}(\epsilon_2,\bR_2,t_2 ) \delta {\cal N}_{\rm T}(\epsilon_1,\bR_1,t_1 ) 
\delta {\cal N}_{\rm T}(\epsilon_0,\bR_0,t_0 )  \rangle = \\
&\langle \delta {\cal N}_{\rm T}(\epsilon_3,\bR_3,t_3 ) \delta {\cal N}_{\rm T}(\epsilon_2,\bR_2,t_2 ) \rangle 
\langle \delta {\cal N}_{\rm T}(\epsilon_1,\bR_1,t_1 )\delta {\cal N}_{\rm T}(\epsilon_0,\bR_0,t_0 )  \rangle \nonumber\\
&+ \delta(\bR_3 -\bR_2)\delta(\bR_2 -\bR_1)\delta(\bR_1 -\bR_0) \delta(\epsilon_3-\epsilon_2) \delta(\epsilon_2-\epsilon_1)
\delta(\epsilon_1-\epsilon_0) \nonumber \\ 
&\times {\cal D}(\epsilon_0,\bR_0)   f_{\rm D}(\epsilon_0- \mu_0)
[1-f_{\rm D}(\epsilon_0- \mu_0)]  [1-2 f_{\rm D}(\epsilon_0- \mu_0)]^2 
\exp[-\gamma(\epsilon_0,\bR_0)(t_3-t_0)]~.\nonumber
 \end{align}
Here, $ {\cal D} ( \epsilon, {\bm R})= \sum_{i=1}^{M_{\rm T}} \delta(\epsilon-\epsilon_i) \delta(\bR-\bR_i) $ 
is the number of trap states per unit volume and energy  at position ${\bm R}$ for energy $\epsilon$.
We assumed that the switching rates $\gamma$ depend only on the trap distance from the graphene layer as ~\cite{mcwhorther_1957,balandin_natnano_2013}
\begin{equation}\label{eq:tunnel}
 \gamma(\epsilon,\bR)=\gamma_0 \exp[-|z-d|/\ell_0]~,
\end{equation}
usually the width of the substrate is $d \sim 100~{\rm nm}$,  and typical orders of magnitude of the tunneling parameters are $\gamma_0  \sim 10^{10}~{\rm s}^{-1}$ and $\ell_0  \sim 1-20~$\AA, respectively~\cite{balandin_natnano_2013,hsu_1970}.
In addition, we assume  that  traps are homogeneously distribuited in the substrate, 
$ {\cal D} ( \epsilon, {\bm R}) \to  {\cal D} ( \epsilon) $. Under these conditions, the deviations of the total number of electrons in the traps  from its equilibrium value is
\begin{equation}\label{eq:dNt}
\delta N_{\rm T} (t) = \int d {\bm r}  \int_0^d dz  \int_{-\Lambda}^{\Lambda} d\epsilon \delta {\cal N}_{\rm T}(\epsilon,\bR,t )~, 
\end{equation}
where $\bR=(\br,z)$, and $\Lambda$ is the cut-off energy.
In the open-circuit limit charge flow between the graphene layer and the metal gate
via the external circuit is not possible. Thus generation-recombination processes in the traps
result in fluctuations of the total number of carriers on the graphene layer 
opposite to fluctuations of the number of electrons in the traps,
i.e. $\delta N (t) = -\delta N_{\rm T} (t) $.
We investigate the critical current noise in  short ballistic GJJs with 
the Dirac-Bogoliubov-de Gennes approach of Ref.~\cite{titov_prb_2006}. 
The resulting critical current, $I_{\rm c}$, depends on the doping level $\mu$ which is, in turn, related to the carrier density. Thus
$\delta \mu \approx (d \mu_0/dn_0) \delta N$, where $\mu_0$ and $n_0$ are respectively the chemical potential and the carrier density in graphene  at  equilibrium. 
Treating  fluctuations of the doping level as an adiabatic perturbation of the Andreev levels, 
critical current fluctuations can be expressed as 
$ \delta I_{\rm c}(t) = I_{\rm c}(t) -\langle I_{\rm c}(t) \rangle$, where
\begin{equation}\label{eq:expgen}
I_{\rm c}(t) \approx  I_{\rm c}(\mu_0) - \frac{d I_{\rm c}}{d \mu_0} \varepsilon_{\rm T}
\delta N_{\rm T} (t)+
 \frac{1}{2}  \frac{d^2 I_{\rm c}}{d \mu_0^2}
 \varepsilon_{\rm T}^2
 [\delta N_{\rm T} (t)]^2~,
\end{equation}
with $\varepsilon_{\rm T}=\frac{e^2}{S C_{\rm Q}}$, $C_{\rm Q}= e^2 \frac{d n_0}{d \mu_0}$ is the quantum capacitance~\cite{ahcn_rmp_2009} and $S$ is the area of the graphene stripe in normal phase.
The power spectrum of the critical current takes the following analytical form
\begin{align}\label{eq:SIc}
&{\cal S}_{ I_{\rm c}} (\omega)= \int_0^\infty \frac{dt}{\pi} \cos(\omega t) \langle \delta  I_{\rm c}(t) \delta  I_{\rm c}(0) \rangle = \frac{\mathcal{A}(\mu, T)}{\omega}=\\
& =  \Bigg[
  \Bigg(  \frac{d I_{\rm c}}{d \mu_0}  \Bigg)^2 F_0 
 - \Bigg(  \frac{d I_{\rm c}}{d \mu_0}  \Bigg) \Bigg(  \frac{d^2 I_{\rm c}}{d \mu_0^2}  \Bigg) \varepsilon_{\rm T} F_1  
 +\Bigg(  \frac{d^2 I_{\rm c}}{d \mu_0^2}  \Bigg)^2   \frac{\varepsilon_{\rm T}^2}{4}  F_2
\Bigg]
\varepsilon_{\rm T}^2 \frac{S \ell_0 }{2 \omega}  {\cal W}(\omega)
~,\nonumber 
\end{align}
where
\begin{equation}
 F_j =  \int_{-\Lambda}^\Lambda d \epsilon  {\cal D}(\epsilon) 
 f_{\rm D}( \epsilon -\mu_0)[1-f_{\rm D}( \epsilon -\mu_0)][1-2f_{\rm D}( \epsilon -\mu_0)]^j~,
\end{equation}
\begin{equation}
 {\cal W}(\omega)=\frac{2}{\pi} \Big[\arctan(e^{d/\ell_0}\omega/\lambda_0)-\arctan(\omega/\lambda_0)\Big]~. 
\end{equation}
In the frequency range $\gamma_0 e^{- d/\ell_0} \ll \omega \ll \gamma_0$, it is easy to verify that $ {\cal W}(\omega) \approx 1$, thus the critical current power spectrum exhibits  flicker noise.
The power spectrum in Eq.~(\ref{eq:SIc}) is the main achievement of this work. This expression has the same form obtained in the fixed gate-voltage limit in Ref.~\cite{pellegrino_commphys_2020}.
The only difference is that in the open-circuit regime $\varepsilon_{\rm T}=e^2/(S C_{\rm Q})$ enters the noise amplitude instead of $\varepsilon_{\rm Q}=e^2/[S (C_{\rm Q}+C_{\rm g})]$, 
where $C_{\rm g}=\epsilon_{\rm r}/(4 \pi d)$ is the geometric capacitance ($\epsilon_{\rm r}$ is the relative dielectric constant of the substrate). For large doping, the quantum capacitance is approximately given by
$C_{\rm Q} \approx \frac{2 e^2  }{\pi \hbar^2 v_{\rm D}^2} |\mu_0| \gg C_{\rm g}$ (where $\vD \sim 10^6~{\rm m/s}$), therefore $\varepsilon_{\rm Q} \approx \varepsilon_{\rm T}$ and the  noise amplitude $\mathcal{A}(\mu,T)$ does 
not depend on the operating condition.
In other words, within the McWorther model, the critical current power spectrum  for large doping does not depend on the thermalization time of the carrier density in graphene.
For zero doping instead, the quantum capacitance is written as  $C_{\rm Q} \approx \frac{2 \ln(2) e^2  }{\pi \hbar^2 v_{\rm D}^2}  k_{\rm B} T \ll C_{\rm g}$,  which leads to $\varepsilon_{\rm Q} \ll \varepsilon_{\rm T}$.
This implies a substantial difference in the critical current spectrum in the two operating conditions. In particular, the temperature dependence is qualitatively different. 
For zero doping it is $d I_{\rm c}/ d \mu_0 |_{\mu_0=0}=0$~\cite{titov_prb_2006}, thus Eq.~(\ref{eq:SIc}) reduces to
\begin{equation}\label{eq:SIc_undoped}
{\cal S}_{ I_{\rm c}} (\omega)  \approx
\Bigg(  \frac{d^2 I_{\rm c}}{d \mu_0^2}  \Bigg)^2   
\frac{\varepsilon_{\rm T}^4 F_2 \ell_0 S}{8\omega} = \frac{\mathcal{A}(0,T)}{\omega}~.
\end{equation}
The temperature dependence is included in  $\varepsilon_{\rm T}$ and  $F_2$.
For low-temperatures  Eq.~(\ref{eq:SIc_undoped}) gives ${\cal A} (0, T) \propto T^{-3}$.
This singular behavior arises from  $\varepsilon_{\rm T}\propto 1/C_{\rm Q}  \propto 1/T$   
and from $F_2 \approx {\cal D}(0) k_{\rm B} T/3 $, valid for a smooth energy dependence of 
the density of trap states ${\cal D}(\epsilon)$.
The singular temperature behaviour of the noise amplitude is a striking consequence of the open-circuit limit.
In the fixed gate voltage regime instead
the energy scale $\varepsilon_{\rm Q}$  tends to the constant value $\varepsilon_{\rm Q}\approx e^2/(S C_{\rm g})$ for low temperatures and the noise amplitude depends  linearly on $T$~\cite{pellegrino_commphys_2020}.

\section{Conclusions}
\label{sec:end}

In this manuscript,  we have studied critical current noise in short ballistic GJJ  based on the phenomenological  McWhorter model~\cite{mcwhorther_1957} in the open-circuit  limit and compared with the fixed gate voltage regime investigated in Ref.~\cite{pellegrino_commphys_2020}.
We have found that in both operating conditions there is a wide frequency range~\cite{balandin_natnano_2013,hsu_1970}, $2 \pi \times  10^{-10}~{\rm s}^{-1}\ll  f \ll 2 \pi \times 10^{10}~{\rm s}^{-1}$, where the critical current spectrum displays the characteristic flicker noise behavior, $ \propto 1/f$.
In the large doping regime, the temperature dependence of the power spectrum turns out not to depend on the
operating regime, 
or equivalently on the charge equilibration time in graphene, as expected. 
At the CNP instead, qualitative differences emerge leading to striking consequences in the low-temperature behavior of the noise amplitude.  In the open circuit limit $\mathcal{A}(T, \mu=0)$ exhibits a singular temperature dependence,  unlike the fixed gate-voltage regime which implies a linear temperature dependence. Our results suggest a viable experimental validation of the phenomenological McWorther noise mechanism
in short-ballistic GJJ by performing critical current noise measurements in open or closed gate voltage circuits.
The ultra-sensitivity of the power spectrum close to the CNP suggests the need for a microscopic modelization of the critical current noise mechanism. The resulting spectrum may have relevant implications on the coherent time evolution of graphene-based gatemons, similarly to qubit~\cite{paladino_rmp_2014,paladino_2003} and qutrit~\cite{falci_2017} realized with conventional Josephson junctions. \\


{\bf Acknowledgements.}
This work has been supported by the Universit\'a degli Studi di Catania, "Linea di intervento 2" of Dipartimento di Fisica e
Astronomia "Ettore Majorana",  Piano di Incentivi per la Ricerca di Ateneo 2016/2018, linea di intervento "Chance" and 2020/2022, project Q-ICT. \\

{\bf Author contributions.} All the authors conceived the work, agreed on the approach to pursue, analysed and discussed the results. F.M.D.P. performed the calculations, E.P. and G.F. supervised the work.


\begin{thebibliography}{}


\bibitem{dean_natnano_2010} 
C.~R. Dean, A.~F. Young,  I. Meric,  C. Lee,  L. Wang,  S. Sorgenfrei,  K. Watanabe,  T. Taniguchi,  P. Kim,  K.~L. Shepard, and J. Hone, 
Nature Nanotech. {\bf 5}, (2010) 722-726.

\bibitem{mayorov_nanolett_2011} 
A.~S. Mayorov,  R.~V. Gorbachev,  S.~V. Morozov,  L. Britnell,  R. Jalil,  L.~A. Ponomarenko,  P. Blake, K.~S. Novoselov,  K. Watanabe,  T. Taniguchi, and  A.~K. Geim,
Nano Lett. {\bf 11}, (2011) 2396-2399.

\bibitem{wang_science_2013}
L. Wang,  I. Meric,  P. Y. Huang,  Q. Gao,  Y. Gao,  
H. Tran,  T. Taniguchi,  K. Watanabe,  L.~M. Campos,  D.~A. Muller,  J. Guo,  P. Kim,  J. Hone,  K.~L. Shepard, and 
C.~R. Dean, 
Science {\bf 342}, (2013) 614.

\bibitem{borzenets_prl_2016} 
I.~V. Borzenets,  F. Amet,  C.~T. Ke,  A.~W. Draelos,  M.~T. Wei,  A. Seredinski,  
K. Watanabe,  T. Taniguchi,  Y. Bomze,  M. Yamamoto,  S. Tarucha, and G. Finkelstein,
Phys. Rev. Lett. {\bf 117}, (2017) 237002.

\bibitem{nanda_nano_2017}
G. Nanda, J.~L. Aguilera-Servin,  P. Rakyta, A. Korm\'anyos, R. Kleine, D. Koelle,  K. Watanabe, T. Taniguchi, L.~M.~K. Vandersypen, and S. Goswami, 
Nano. Lett. {\bf 17}, (2017) 3396-3401.

\bibitem{park_prl_2018} 
J. Park,  J. H. Lee,  G.-H. Lee,  Y. Takane,  K.-I. Imura,  T. Taniguchi,  K. Watanabe, and H.-J. Lee,
Phys. Rev. Lett. {\bf 120}, (2018) 077701.

\bibitem{benshalom_natph_2015} 
M. Ben Shalom,  M.~J. Zhu,  V. I. Fal'ko,  A. Mishchenko,  A.~V. Kretinin,  K.~S. Novoselov, C.~R. Woods,  K. Watanabe,  T. Taniguchi,  A.~K. Geim, and J.~R. Prance,
Nature Phys. {\bf 12}, (2016) 318-322.

\bibitem{calado_natnano_2015}
V.~E. Calado,  S. Goswami,  G. Nanda,  M. Diez,  A.~R. Akhmerov,  K. Watanabe,  T. Taniguchi, 
 T.~M. Klapwijk, and L.~M.~K. Vandersypen, 
Nature Nano. {\bf 10}, (2015) 761.

\bibitem{english_prb_2016} 
C.~D. English,  D.~R. Hamilton,  C. Chialvo,  I.~C. Moraru,  N. Mason, and D.~J. Van Harlingen, 
Phys. Rev. B {\bf 94}, (2016) 115435.

\bibitem{titov_prb_2006} 
M. Titov, and  C.~W.~J. Beenakker, Phys. Rev. B {\bf 74}(R), (2006) 041401.

\bibitem{wang_natnanotech_2018}
J.~I-Jan Wang,  D. Rodan-Legrain,  L. Bretheau,  D.~L. Campbell,  B. Kannan,  
D. Kim, M. Kjaergaard,  P. Krantz,  G.~O. Samach,  F. Yan,  J.~L. Yoder,  K. Watanabe,  T. Taniguchi,  T.~P. Orlando, 
 S. Gustavsson,  P. Jarillo-Herrero, and W.~D. Oliver,
Nature Nanotech. {\bf 14}, (2018) 120-125.

\bibitem{pellegrino_proceedings_2019}
F.~M.~D. Pellegrino,  G. Falci, and E. Paladino,  Proceedings {\bf 12}, (2019) 33. 

\bibitem{mcwhorther_1957}
A.~L. McWhorter, {\it Semiconductor Surface Physics},  R. H. Kingston (University of Philadelphia Press, Philadelphia, PA, 1957), p. 207.

\bibitem{hooge_rps_1981} 
F.~N. Hooge,  T.~G.~M. Kleinpenning, and  L.~K.~J. Vandamme, 
Rep. Prog. Phys. {\bf 44}, (1981) 479-532.

\bibitem{balandin_natnano_2013} 
A.~A. Balandin, 
 Nature Nano. {\bf 8}, (2013) 549.

\bibitem{paladino_rmp_2014}
E. Paladino,  M.~Y. Galperin,  G. Falci, and B.~L. Altshuler, 
 Rev. Mod. Phys. {\bf 86}, (2014) 361.

\bibitem{darrigo_2008} 
A. D'Arrigo, A. Mastellone, E.  Paladino, and G. Falci, New J. Phys. {\bf 10}, (2008) 115006.

\bibitem{paladino_2010} E. Paladino, A. Mastellone, A. D'Arrigo, G. Falci,  Phys. Rev. B {\bf 81}, (2010) 052502.

\bibitem{pellegrino_commphys_2020}
F.~M.~D. Pellegrino, G. Falci, and E. Paladino, Commun. Phys. {\bf 3}, (2020) 6.

\bibitem{kogan_book} 
S. Kogan, {\it Electronic noise and fluctuations in solids}, (Cambridge University Press: Cambridge, UK, 1996), p. 118.


\bibitem{pellegrino_jstat_2019}
F.~M.~D. Pellegrino, G. Falci, and E. Paladino, J. Stat. Mech. {\bf 2019}, (2019) 094015.

\bibitem{ahcn_rmp_2009}
A.~H. Castro Neto,  F. Guinea,  N. M. R. Peres,  K. S. Novoselov, and A. K. Geim,
{\it Rev. Mod. Phys.} {\bf 81}, (2009) 109.


\bibitem{hsu_1970}
S.~T. Hsu,  Solid-State Electronics {\bf 13}, (1970) 843-855.


\bibitem{paladino_2003} E. Paladino, L. Faoro, and G. Falci, Adv. in Sol. State Phys. {\bf 43},
(2003) 747.


\bibitem{falci_2017} 
G. Falci, P.~G. Di Stefano, A. Ridolfo, A. D'Arrigo, G.~S. Paraoanu, and E. Paladino
Fortschr. Phys. {\bf 65}, (2017)  1600077.


\end{thebibliography}
\end{document}